# Utilizing Noise Addition for Data Privacy, an Overview

Kato Mivule
Computer Science Department
Bowie State University
14000 Jericho Park Road Bowie, MD 20715
Mivulek0220@students.bowiestate.edu

**Abstract** – *The internet is increasingly becoming a standard for both the production and consumption of data while at the same time cyber-crime involving the theft of private data is growing. Therefore in efforts to securely transact in data, privacy and security concerns must be taken into account to ensure that the confidentiality of individuals and entities involved is not compromised, and that the data published is compliant to privacy laws. In this paper, we take a look at noise addition as one of the data privacy providing techniques. Our endeavor in this overview is to give a foundational perspective on noise addition data privacy techniques, provide statistical consideration for noise addition techniques and look at the current state of the art in the field, while outlining future areas of research.*

## 1. Introduction

Large data collection organizations such as the Census Bureau often release statistics to the public in the form of statistical databases, often transformed to some extent, omitting sensitive information such as personal identifying information (PII). Researchers have shown that with such publicly released statistical databases in conjunction with supplemental data, adversaries are able to launch inference attacks and reconstruct identities of individuals or an entity's sensitive information [1]. Therefore while data de-identification is essential, it should be taken as an initial step in the process of privacy preserving data publishing but other methods such as noise addition should strongly be considered after PII has been removed from data sets to ensure greater levels of confidentiality [1] [2]. A generalized data privacy procedure would involve both data de-identification and perturbation as shown in *Figure 1*.

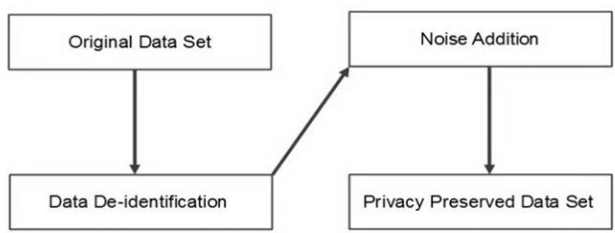

***Figure 1:*** *Generalized Data Privacy with Noise Addition*

## 2. Background

In this section we take a look at some of the terms used in the noise addition procedure. *Data Privacy and Confidentiality* is the protection of an entity or an individual against illegitimate information revelation. [1]. *Data Security* is concerned with legitimate accessibility of data [2]. *Data de-identification* is the removal of personally identifiable information (PII) from a data set [3] [4]. *Data de-identification process* also referred to as *data anonymization, data sanitization,* and *statistical disclosure control* (SDC), is a process in which PII attributes are excluded or denatured to such an extent that when the data is made public, a person's identity, or an entity's sensitive data, cannot be reconstructed [5] [6]. *Statistical disclosure control* methods are classified as *non-perturbative* and *perturbative*, with the former being a procedure in which original data is not denatured, while with the latter, original data is denatured before publication to provide confidentiality [1]. Therefore de-identification of data ensures to some extent that sensitive and personal data does not suffer from *inference* and *reconstruction attacks*, which are methods of attack in which isolated pieces of data are used to infer a supposition about a person or an entity [7].

*Data utility verses privacy* is how useful a published dataset is to the consumer of that publicized dataset. In most instances, when publishers of large data sets do so, they ensure that PII is removed and data is distorted by noise addition techniques. However, in doing so, the original data suffers loss of some of its statistical properties even while confidentiality is granted, thus making the dataset almost meaningless to the user of the published dataset. Therefore a balance between privacy and utility needs is always sought [24] [25] [26]. Data privacy scholars have noted that achieving optimal data privacy while not shrinking data utility is an ongoing NP-hard task [27]. *Statistical databases* are non-changing data sets often published in aggregated format [28]. While data de-identification will ensure the removal of PII attributes, it has been deemed a novice method by researchers; the remaining *sanitized* data set could still be compromised and used to reconstruct an individual's identity or an entity's sensitive data [1] [2]. Therefore the remaining confidential attributes that contain sensitive information for example salary, student's GPA, need to be transformed to such an extent that they cannot be linked with outside information in an inference attack. It is in this context that we focus on noise addition as a perturbation methodology that seeks to transform numerical attributes to grant confidentiality.

## 3. Related work

With an increasing interest in data privacy and security research, a number of surveys have been done articulating the progress and state of the art in the data privacy and security research field. In their survey on data privacy and security, Santos et al., present an overview on state of the art in data security techniques, placing emphasis on data security solutions for data warehousing [40]. Furthermore, in their overview, Matthews and Harel, offer a more comprehensive summary of current statistical disclosure limitation techniques, noting that that the balance between privacy and utility is still being sought with data privacy enhancing techniques [41]. Additionally Joshi and Kuo, offer an outline of state of the art data privacy techniques in Online Social Networks, in which they note how a balance is always pursued between privacy requirements for users and using private data for advertisements [42]. Yet still, in their review, Ying-hua et al., take a closer look at the current data privacy preserving techniques in data mining, providing advantages and disadvantages of various data privacy procedures [43]. While a number of current overviews on data privacy focus on the general data privacy enhancing techniques, in this paper, we focus on noise addition methods while providing statistical considerations for data perturbation.

## 4. Noise Addition

In this section, we take a look at noise addition perturbation methods that transform confidential attributes by adding noise to provide confidentiality. Noise addition works by adding or multiplying a stochastic or randomized number to confidential quantitative attributes. The stochastic value is chosen from a normal distribution with zero mean and a diminutive standard deviation [10] [11].

### 4.1. Additive Noise

Work on additive noise was first publicized by Kim [12] with the general expression that

$$Z = X + \varepsilon \qquad (1)$$

Where $Z$ is the transformed data point, $X$ is the original data point and $\varepsilon$ is the random variable (noise) with a distribution $e \sim N(0, \sigma^2)$. This is then added to $X$. The $X$ is then replaced with the $Z$ for the data set to be published.[13] *With stochastic noise,* random data is added to confidential attributes to conceal the distinguishing values, an example includes increasing a student's GPA by a diminutive percentage, say from 3.45 to 3.65 GPA [14]. In their work on additive noise, Domingo-Ferrer et al., outline that in additive noise, also referred to as *white noise*, concealment by additive noise anticipates that the variable of measurements $x_j$ of the original data set $X_j$ is continuously replaced by the variable,

$$z_j = x_j + j \qquad (2)$$

Where $j$ is the variable of normally distributed noise acquired from a random variable: $\varepsilon_j \sim N(0, \sigma_j^2)$, such that
$Cov(\varepsilon_t, \varepsilon_l)$, for all $t! = l$ thus the method preserves the mean and covariance. [20] Therefore additive noise can be expressed in a simple format as follows [21]:

$$Z = X + \varepsilon \qquad (3)$$

$Z$ is masked data value to be published, after the transformation $X + \varepsilon$. $X$ is the original unmasked data value in the raw data set. $\varepsilon$ (epsilon) is the random variable (noise) added to $X$, whose distribution is $\varepsilon \sim N(0, \sigma^2)$. Ciriani et al., note that additive noise also known as *uncorrelated noise*, preserves the mean and covariance of the original data but the correlation coefficients and variances are not sustained. Another variation of additive noise is *correlated additive noise* that keeps the mean and allows the sustenance of correlation coefficients in the original data [22].

### 4.2. Multiplicative Noise

Multiplicative noise is another type of stochastic noise outlined by Kim and Winkler [23] in which they describe that multiplicative noise is rendered by generating random numbers with a mean = *1*, which then are used as noise and multiplied to the original data set. Each data element is multiplied by a random number with a short Gaussian distribution, with mean = *1* and a small variance:

$$Y_j = X_j \varepsilon_j \qquad (4)$$

Where *Y* is the perturbed data; *X* is the original data; *E* is the generated random variable (noise) with a normal distribution with mean *μ* and variance *σ* [23].

### 4.3 Logarithmic multiplicative noise

Kim and Winkler [23] describe another variation of multiplicative noise, in which a logarithmic alteration is taken on the original data:

$$Y_j = lnX_j \qquad (5)$$

The random number (noise) is then generated and then added to the altered data [23]:

$$Z_j = Y_j + \varepsilon_j \qquad (6)$$

Where *X* is the original data; *Y* is the logarithmic altered data; *Z* is the logarithmic altered data with noise added to it; $e^x$ is the exponential function used to calculate the antilog.

### 4.4. Differential Privacy

In this section, we take a look at Differential privacy, a current state of the art data perturbation method that utilizes Laplace noise addition techniques and was proposed by Dwork (2006). Differential privacy is the latest state-of-the-art methodology in data privacy that enforces confidentiality by returning perturbed aggregated query results from databases, such that users of the databases cannot discern if particular data item has been altered or not. This means that with the perturbed results of the query, an attacker cannot derive information about any data item in the database [33]. The database in this case is a collection of rows that represent each individual entity we seek to provide concealment. [34] According to Dwork (2008), two databases $D_1$ and $D_2$ are considered identical or similar, if they differ or disagree in only one element or row that is $D_1 \Delta D_2 = 1$. Therefore, a procedure $q_n$ that grants confidentiality, satisfies *ε-differential privacy* if the result to any same query run on database $D_1$ and again run on database $D_2$ should probabilistically be similar, and as long as those results satisfy the following requirement: [36]

$$\frac{P[q_n(D_1) \in R]}{P[q_n(D_2) \in R]} \leq e^\varepsilon \quad (7)$$

Where $D_1$ and $D_2$ are the two databases
- $P$ is the probability of the perturbed query results $D_1$ and $D_2$ respectively.
- $q_n()$ is the privacy granting procedure (perturbation).
- $q_n(D_1)$ is the privacy granting procedure on query results from database $D_1$.
- $q_n(D_2)$ is the privacy granting procedure on query results from database $D_2$.
- $R$ is the perturbed query results from the databases $D_1$ and $D_2$ respectively.
- $e^\varepsilon$ is the exponential *ε* epsilon value.

Therefore to satisfy differential privacy, the probability of the perturbed query results $D_1$ divided by the probability of the perturbed query results $D_2$ should be less or equal to an exponential *ε* epsilon value. That is to say, if we run the same query on database $D_1$, and then run the same query again on database $D_2$, our query results should probabilistically be similar. If the condition can be mitigated in the presence or absence of the most influential observation for a particular query, then this condition will also be mitigated for any other observation. The consequence of the most dominant observation for a given query is given by $\Delta f$ and assessed in the following way:

$$\Delta f = Max|f(D_1) - f(D_2)| \quad (8)$$

For all possible realizations of $D_1$ and $D_2$, Where $f(D_1)$ and $f(D_2)$ represent the true responses to the query from $D_1$ and $D_2$ [33] [34] [35] [36]. According to Dwork (2006), the results to a query are presented as noise in the following way:

$$f(x) + Laplace(0, b) \quad (9)$$

Where $b$ is defined as follows for Laplace noise:

$$b = \frac{\Delta f}{\varepsilon} \quad (10)$$

$X$ represents a particular realization of the database, while $f(x)$ represents the true response to the query, the response would satisfy *ε-differential privacy*. The $\Delta f$ must look at all possible realizations of $D_1$ and $D_2$ [33] [34] [35] [36] [37]. We could take an example in which we query the GPA of students at Bowie State University. If our Min GPA in the database is 2.0, for smallest possible GPA, and our Max GPA is 4.0 for largest possible GPA, we then calculate $\Delta f$ as 2.0. We choose a small *ε* value of 0.01. The parameter $b$ of the Laplace noise is set to $\Delta f / \varepsilon = 2.0/0.01 = 200$. Thus we have Laplace (0, 200) noise distribution. Therefore the unperturbed results of the query + Noise from Laplace (0, 200) = Perturbed query results satisfying *ε-differential privacy*. [34] It has been noted by researchers that a smaller *ε* epsilon value creates greater privacy by the procedure. However, utility risks degeneration with a much smaller *ε* epsilon value [38]. For example, *ε* at 0.0001, will give *b* as 20000, Laplace (0, 20000) noise distribution.

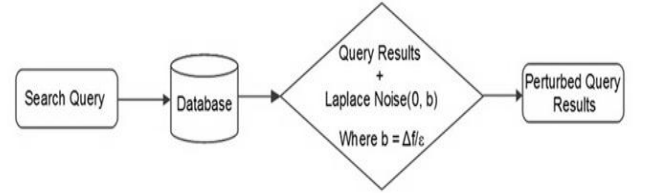

*Figure 2:* A general Differential Privacy satisfying procedure.

General steps for differential privacy shown in *Figure 2:*
- Run query on database
- Calculate the most influential observation
- Calculate the Laplace noise distribution
- Add Laplace noise distribution to the query results
- Publish perturbed query results.

### 4.5. Differential privacy pros and cons
Differential privacy grants across-the-board privacy, and easy to implement with SQL for aggregated data publication [39]. However, utility is a challenge as statistical properties change with a much smaller *ε*, as Laplace noise addition takes into account the outliers and most influential observation. [38] More noise to the data at the level of the most influential observation only renders the data useless thus balance between privacy and utility still a challenge [34] [37].

### 4.6. Statistical background for Noise addition
In this section, we take a look at statistical considerations for data perturbation utilizing noise addition. With noise addition, transformed data has to keep the same statistical properties as the original data. Therefore consideration has to be made for statistical characteristics such as normal distribution, mean, variance, standard deviation, covariance, and correlation for both original and perturbed data sets.

*The Mean μ,* is the average of values after their total sum has been taken. In this case we would look at the summation of values then we divide them by the *n*, the quantity of values; the mathematical statement then for *the Mean μ,* is straight forward: [16]

$$\mu = \frac{1}{n} \sum_{k=0}^{n} x_i \quad (11)$$

*The Normal Distribution*, also known as the Gaussian distribution, used in calculating the noise addition, is a bell shaped continuous probability distribution used as an estimation to depict real-valued stochastic variables that agglomerate around a single mean. The formula for normal distribution is as follows:[15]

$$f(x) = \frac{1}{\sqrt{(2\pi\sigma^2)}} \times e^{-((x-\mu)^2/2\sigma^2)} \quad (12)$$

The parameter $\mu$ represents the mean, the point of the peak in the bell curve, while the parameter $\sigma^2$ representing the variance, the width of the distribution. The annotation $N(\mu, \sigma^2)$ represents a normal distribution with mean $\mu$ and variance $\sigma^2$. Therefore $X \sim N(\mu, \sigma^2)$ is representative of X distributed $N(\mu, \sigma^2)$. The distribution with $\mu = 0$ and $\sigma^2 = 1$ is cited to as the standard normal.

*The Variance $\sigma^2$*, in noise addition, is a measure of how data distributes itself in approximation to the mean value. The expression for variance is given by: [17]

$$\sigma^2 = \frac{\sum (X-\mu)^2}{N} \quad (12)$$

Where $\sigma^2$ is the variance, $\mu$ is the mean, X being the single data values, N as the number of values, and $\sum (X-\mu)^2$ as the summing up of all data values X minus the mean $\mu$ squared.

*The Standard Deviation, σ,* is a measure of how distributed data is from the normal, thus we would say standard deviation is how data points are deviated from the mean. The mathematical expression is simply the square root of the variance $\sigma^2$: [18]

$$\sigma = \sqrt{\sigma^2} \quad (13)$$

*Covariance:* With noise addition, the measurement of how affiliated original data and perturbed data are, is crucial. Covariance, *Cov(X, Y)*, is a calculation of how affiliated the deviations between the data points *X* and *Y* are. If the covariance is positive, then the *X* and *Y* data points' inclination is to increase together, else if the covariance is negative, then the tendency is that for the two data points *X* and *Y*, one lessens while the other gains. However, if the covariance is zero, then this would signal that the data points are each autonomous. The expression for covariance is given as follows: [19]

$$Cov\ xy = \frac{1}{N} \sum x_i y_i - \overline{xy} \quad (14)$$

*Correlation $r_{xy}$* also known as the Pearson product, calculates the capability and inclination of an additive or linear relation between two data points. The correlation $r_{xy}$ is dimensionless, autonomous of the parts in which the data points *x* and *y* are calculated [19]. If $r_{xy}$ is = *-1*, then $r_{xy}$ indicates a negative linear relation between the data points *x* and *y*. If $r_{xy} = 0$, then the linear relation between the two data points *x* and *y* does not exist, however, a regular nonlinear relation might exist. If $r_{xy} = +1$, then there is a strong linear relation between *x* and *y*. The expression used for correlation is: [19]

$$Correlation = r_{xy} = \frac{Cov\ xy}{\sigma_x \sigma_y} \quad (15)$$

### 4.7. Signal Noise Ratio (SNR)

In this section, we take a look at SNR in relation to data perturbation using noise addition, with the aim that SNR could be employed to achieve optimal data utility while preserving privacy, by measuring how much noise we need to optimally obfuscate data. In electronic signals, SNR is used to calculate a signal tainted by noise by approximating the signal power to noise power ratio, basically the ratio of the power of the signal without noise over the power of the noise.

$$SNR = \frac{Signal\ Varince}{Noise\ Variance} \quad (16)$$

With data perturbation, we could further borrow from the definition of SNR employed in Image Processing were the ratio of mean to standard deviation of a signal is used, and typically SNR is computed as the ratio of the mean pixel value to the standard deviation of the pixel values in a certain vicinity [29] [30].

$$SNR = \frac{\mu}{\sigma} \quad (17)$$

The parameter $\mu$ in this case represents the mean of the signal and the parameter $\sigma$ as the standard deviation of the noise. A presumed threshold for *SNR* in image processing is based on the *Rose Criterion* which stipulates that an *SNR* of 5 is desirable in order to differentiate image details with 100 per cent confidence. Therefore, an *SNR* of less than 5 per cent will result in less than 100 per cent confidence in recognizing particulars of an image [31].

### 5. Illustration

In this section, we provide an example of data perturbation with noise addition for illustrative purposes. We follow a simple algorithm in implementing noise addition perturbation methodology to provide confidentiality in a published data set. The first step is the de-identification of the data set by the removal of PII, after which we apply noise addition. In our implementation, we created a data set of 10 records for illustrative purposes and then applied the algorithm below. The original data set contained PII, we de-identified the original data set, after which we applied additive noise to the numerical attributes, and we then plotted the results in a graph, comparing the statistical properties of the original and perturbed data.

*Steps for De-identification and Noise Addition*
1. For all values of the data set to be published,
2. Do data de-identification
    2.1. Find PII
    2.2 Remove PII
3. For remaining data void of PII to be published,
    3.1. Find quantitative attributes in the data set
    3.2. Apply additive noise to the quantitative data values
4. Publish data set

**5.1. Results of Illustration**

| First Name | Last Name | SSN | Age | Major | GPA | Zip code | State | Scholarship Amount | Gender |
|---|---|---|---|---|---|---|---|---|---|
| John | Artist | Xxx-xx-xxx9 | 23 | Computer Science | 4.00 | 21071 | MA | 30000.00 | M |
| Peter | Chemist | Xxx-xx-xx10 | 33 | Biology | 3.35 | 31072 | MD | 50000.00 | M |
| Evan | Biologist | Xxx-xx-xx11 | 32 | Chemistry | 2.19 | 21073 | MA | 25000.00 | F |
| Joy | Music | Xxx-xx-xx12 | 45 | History | 2.99 | 21074 | MD | 67000.00 | F |
| Eva | Pictures | Xxx-xx-xx13 | 23 | Chemistry | 3.67 | 21075 | PA | 78000.00 | F |
| Sandra | Hollywood | Xxx-xx-xx14 | 21 | Art | 3.65 | 21076 | MD | 90888.00 | F |
| Okello | Oscars | Xxx-xx-xx15 | 25 | Music | 4.00 | 21077 | LA | 90000.00 | M |
| Mukisa | Grammys | Xxx-xx-xx16 | 30 | Computer Science | 2.79 | 21078 | PA | 10000.00 | M |
| Jacinta | Historian | Xxx-xx-xx17 | 32 | Biology | 3.00 | 21079 | CA | 7000.00 | F |
| Bosco | Activist | Xxx-xx-xx18 | 37 | Art | 2.98 | 21080 | MO | 11000.00 | M |

**Table 1:** *Original Data Set (All data for illustrative purposes).*

| Age | Major | GPA | State | Scholarship Amount |
|---|---|---|---|---|
| 23 | Computer Science | 3.33 | MA | 30000.00 |
| 33 | Biology | 3.35 | MD | 50000.00 |
| 32 | Chemistry | 2.19 | MA | 25000.00 |
| 45 | | 2.99 | MD | 67000.00 |
| 23 | Chemistry | 3.35 | PA | 78000.00 |
| 21 | Art | 2.19 | MD | 90888.00 |
| 25 | | 3.11 | LA | 90000.00 |
| 30 | Computer Science | 2.99 | PA | 10000.00 |
| 32 | Biology | 3.00 | | 7000.00 |
| 37 | Art | 3.00 | | 11000.00 |

**Table 2**: *Result after de-identification on original data.*

| Scholarship Amount | Normal D of Origin Scholarship | Perturbed Scholarship Amount | Normal D of Perturbed Scholarship |
|---|---|---|---|
| 30000.00 | 0.7602 | 37296.88 | 0.7788 |
| 50000.00 | 0.9859 | 52087.4 | 0.9555 |
| 25000.00 | 0.6312 | 29403.32 | 0.6066 |
| 67000.00 | 0.9997 | 72389.89 | 0.9986 |
| 78000.00 | 1.0000 | 84639.16 | 0.9999 |
| 90888.00 | 1.0000 | 97116.52 | 1.0000 |
| 90000.00 | 1.0000 | 91554.68 | 1.0000 |
| 10000.00 | 0.2172 | 18977.3 | 0.3494 |
| 7000.00 | 0.1575 | 10455.81 | 0.1777 |
| 11000.00 | 0.2398 | 17932.86 | 0.3253 |

**Table 4:** *Results of the Normal Distribution of Original Perturbed Scholarship Amount.*

| | | Random Noise Between 1000 and 9000 | |
|---|---|---|---|
| | Scholarship Amount | Random Noise | Perturbed Scholarship Amount |
| | 30000.00 | 7296.88 | 37296.88 |
| | 50000.00 | 2087.4 | 52087.4 |
| | 25000.00 | 4403.32 | 29403.32 |
| | 67000.00 | 5389.89 | 72389.89 |
| | 78000.00 | 6639.16 | 84639.16 |
| | 90888.00 | 6228.52 | 97116.52 |
| | 90000.00 | 1554.68 | 91554.68 |
| | 10000.00 | 8977.3 | 18977.3 |
| | 7000.00 | 3455.81 | 10455.81 |
| | 11000.00 | 6932.86 | 17932.86 |
| Mean | 20500 | | 27614.87 |
| Standard Deviation | 13435.0288425444 | | 13692.4298530319 |
| Variance | 180500000 | | 187482635.2802 |

**Table 3:** *Random noise between 1000 and 9000 added to Scholarship attribute.*

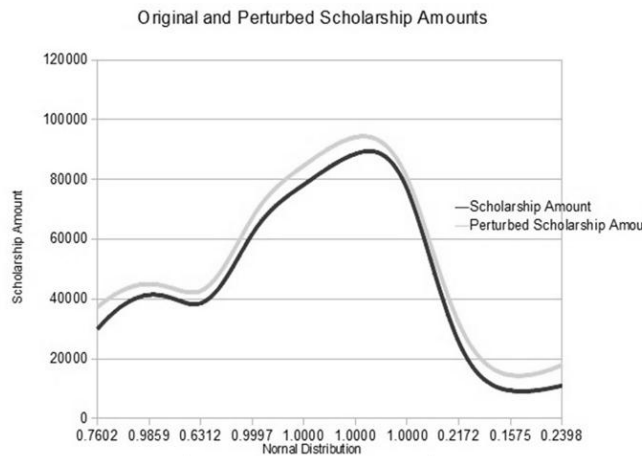

*Figure 3*: Results of the normal distribution of original and perturbed scholarship amount

*Covariance* between Original Scholarship Data set and Perturbed Scholarship Data set = *1055854875.465*. Since Covariance is positive, it shows that the two data sets move together in the same direction. *Correlation* between Original Scholarship Data set and Perturbed Scholarship Data set = *0.999*. Since Correlation is a strong positive, it shows a relationship between the two data sets, increasing and decreasing together.

## 6. Conclusion

We have taken a look at data perturbation utilizing noise addition as a methodology used to provide privacy for published data sets. We also took a look the statistical considerations when utilizing noise addition. We provided an illustrative example showing that de-identification of data when done in concert with noise addition would add more to the privacy of published data sets while maintaining the statistical properties of the original data set. However, generating perturbed data sets that are statistically close to the original data sets is still a challenge as consideration has to be made for the tradeoff between utility and privacy; the more close the perturbed data is to the original, the less confidential that data set becomes, and the more distant the perturbed data set is from the original, the more secure but then, utility of the data set might be lost when the statistical characteristics of the origin data set are lost. Noise generation certainly affects the level of perturbation on the original data set. Yet still, striking the right balance between privacy and utility remains a factor. While state of the art data perturbation techniques such as differential privacy provide hope for achieving greater confidentiality, achieving optimal data privacy while not shrinking data utility is an ongoing NP-hard task. Therefore more research needs to be done on how optimal privacy could be achieved without degrading data utility. Another area of research is how noise addition techniques could be optimally applied in the cloud and mobile computing arena, given the ubiquitous computing era.

Databases, vol. 3050, Springer Berlin / Heidelberg, 2004, p. 519.